\documentclass[preprint,aps,floatfix,amsmath]{revtex4}
\usepackage{epsfig}

\usepackage{longtable}
\usepackage{bm}

\begin{document}
\title{Experimental demonstration of multi-watt CW supercontinuum tailoring in photonic crystal fibers}
\author{A. Kudlinski\footnote{Electronic mail address: \tt alexandre.kudlinski@univ-lille1.fr}, G. Bouwmans,
Y. Quiquempois and A. Mussot}

\affiliation{Universit\'e des Sciences et Technologies de Lille,\\
IRCICA, FR CNRS 3024,\\ Laboratoire PhLAM, UMR CNRS 8523,\\ 59655
Villeneuve d'Ascq Cedex, France}

\begin{abstract}
We demonstrate experimentally that the spectral broadening of CW
supercontinuum can be controlled by using photonic crystal fibers
with two zero-dispersion wavelengths pumped by an Yb fiber laser at
1064~nm. The spectrum is bounded by two dispersive waves whose
spectral location depends on the two zero-dispersion wavelengths of
the fiber. The bandwidth of the generated spectrum and the spectral
power density may thus be tailored for particular applications, such
as high-resolution optical coherence tomography or optical
spectroscopy.
\end{abstract}
\maketitle

The continuous-wave (CW) pumping regime has been recently proposed
as an interesting alternative to pulsed lasers for the generation of
strong supercontinuum (SC)
\cite{taylor03,miguel03,abeeluck04,abeeluck05,travers05}. Sources
based on CW SC are characterized by a substantial lower intensity
noise, a lower coherence length, a higher stability and a higher
spectral power density than their pulsed counterparts. Additionally,
photonic crystal fibers (PCFs) can be spliced to recently developed
CW fiber lasers, preserving the all-fiber format. These features are
of particular interest for ultrahigh resolution optical coherence
tomography for instance. SC spectra generated with a PCF pumped by a
CW Yb fiber laser were reported in Refs.~\cite{taylor03} and
\cite{travers05}. About 700~nm-wide spectra with high spectral power
densities (in the order of 10~mW/nm) were obtained. But one drawback
of these configurations is that the bandwidth of the SC spectrum is
not easily adjustable to the one required for a particular
application. As a consequence the pump power budget is not optimized
if the application requires a given spectral power density over a
particular spectral range, as this is actually the case for many
potential applications. Quite recently, it has however been
numerically demonstrated that a simple scheme based on a PCF with
two-zero dispersion wavelengths allows a tailoring of the SC
spectrum extent, and consequently of the spectral power density
\cite{arnaud07}. These numerical results are of great potential
interest for the reasons stated before. In this paper, we provide an
experimental proof-of-principle of this novel technique by pumping
PCFs with a CW Yb fiber laser. We experimentally show that the
generated SC spectrum can be tailored through a suitable design of
the group velocity dispersion (GVD) of the fiber, and in particular
of the position of its two ZDWs, as predicted numerically
\cite{arnaud07}. We report the generation of 200~nm and 500~nm wide
spectra in two slightly different PCFs with two zero-dispersion
wavelengths (ZDWs). The SC is achieved by pumping in the anomalous
dispersion region, just between the two ZDWs. We showed that a
slight modification of the dispersion curve allows the control of
the SC extension and consequently permits to optimize the pump power
budget for particular applications.

For the sake of simplicity and clarity, we recall here only on the
basic mechanisms involved in SC generation in CW pumping regime. A
detailed theoretical analysis of the SC formation is given in
Ref.~\cite{arnaud07}. In the first stage, modulational instability
converts the CW field into a train of quasi-solitonic pulses of a
few picoseconds duration \cite{agrawal}. These pulses experience
higher-order linear and nonlinear effects so that they become
unstable. Consequently, they shed energy away to dispersive waves
(DWs) which are phase-matched with the solitonic waves
\cite{cristiani04,akhmediev95}. From this phase-matching condition
that depends on the GVD and on the nonlinear phase mismatch of the
solitonic waves \cite{akhmediev95}, one can estimate the spectral
position of the DWs. On the contrary to fibers with a single ZDW
where only one DW is phase-matched with one solitonic pulse, in fibers
with two ZDWs there are two wavelengths for which the phase-matching
condition is satisfied. Consequently, two DWs are generated: one is
located on the blue side of the lower ZDW and the other one is
located on the red side of the higher ZDW. Additionally the
stimulated Raman scattering shifts the solitons towards longer
wavelengths \cite{agrawal}. This so-called soliton self-frequency
shift stops just below the second ZDW because a balance is achieved
between the red shift due to SRS and the blue shift due to spectral
recoil when the dispersion slope of the fiber is negative
\cite{skryabin03}. It is worth noting that even more complex
nonlinear interactions should occur during the propagation of such
strong pulses in a low dispersion region, such as soliton collisions
that can reinforce the frequency shift towards long wavelengths
\cite{frosz}. Providing that the pump wavelength is located between
both ZDWs in a low anomalous dispersion regime, one can simply
summarize that the SC extension is roughly limited by the spectral
separation between the two DWs.

In order to experimentally demonstrate the numerical results
predicted in Ref.~\cite{arnaud07}, we designed two PCFs with two
ZDWs from part to part of the pump wavelength at 1064~nm, and a low
anomalous dispersion region at this wavelength. This can be achieved
for a relatively small hole-to-hole spacing $\Lambda$ in the order
of 1.6~$\mu$m, and a $d/\Lambda$ value of about 0.4, with $d$ being
the hole diameter \cite{arnaud07,tse06}. For a reasonable number of
seven air-hole rings, such a small pitch leads to relatively
important confinement losses around 1500~nm, where the SC is
expected to extend. We numerically evaluated these losses with a
finite-elements method (FEM), and we found that they were in the
order of 1000~dB/km at 1500~nm. To overcome this problem, we
replaced the seventh air-hole period with a ring of larger air holes
($d/\Lambda = 0.8$), as can be seen in the scanning electron
micrograph (SEM) represented in the inset of Fig.~\ref{fig1}.
Confinement losses were thus reduced to about 1~dB/km at 1500~nm,
which is an acceptable value for experiments performed in tens of
meters long fibers, as required for CW SC generation. Another way to
reduce confinement losses without adding such an air-clad structure
would be to use a higher number of uniform air-hole rings (typically
ten or eleven). However, this would make the fabrication process a
lot trickier since this would approximately double the total number
of holes in the cladding.

We fabricated two PCFs labeled fiber A and fiber B in what follows.
They were designed to exhibit a GVD curve similar to that of PCF1
and PCF2 in Ref.~\cite{arnaud07}. The GVD curves of the fabricated
fibers were calculated from high resolution SEMs with a FEM package.
They are represented in Fig.~\ref{fig1} together with an example of
SEM for fiber B in inset. They both exhibit a convex shape with two
ZDWs located on each side of the pump wavelength (represented by the
vertical dashed line in Fig.~\ref{fig1}). The basic characteristics
of the fabricated PCFs are also summarized in Table~\ref{tab1} for
the sake of clarity. It can be seen from this table that the
parameters of fibers A and B are very close to those of PCF1 and
PCF2 of Ref.~\cite{arnaud07} respectively. The nonlinear coefficient
$\gamma$ was deduced from the computed effective area at 1064~nm, by
taking the typical value of $n_2$ =
2.6$\times$10$^{-20}$~m$^2$.W$^{-1}$ for the nonlinear refractive
index of silica \cite{agrawal}. The nonlinear coefficient of both
PCFs was in the order of 25~W$^{-1}$.km$^{-1}$ at 1064~nm. Note that
this value is about twice higher than the nonlinear coefficient of
PCFs with a single ZDW around 1064~nm usually designed for SC
generation \cite{wadsworth04}. Since the two ZDWs of fiber A are
closer than the ones of fiber B, a broader SC is expected in fiber B
according to the theoretical study summarized above and detailed in
Ref.~\cite{arnaud07}.

The PCFs were pumped with a linearly polarized CW Yb fiber laser
(IPG) delivering 20~W at 1064~nm with a full width half maximum of
0.5~nm. The setup is displayed in Fig.~\ref{fig2}(a). The collimated
beam with a 5~mm diameter (at $1/e^2$) was sent through an afocal
setup made of two IR-coated lenses in order to reduce the spot size
by a factor of 2. About 70 \% of the pump power was launched into
the fibers by using appropriate aspherical lens. However, even
though most of the pump power is coupled into the core of the PCFs,
there is a small amount of pump power launched and guided in the
inner cladding because of the last ring of holes (larger air holes
with $d/\Lambda = 0.8$). Consequently the total power of 14~W
launched into the fibers does not contribute completely to the SC
generation. The output light was butt-coupled to a standard
single-mode pigtail in order to record the spectra with an optical
spectrum analyzer (OSA). Figure~\ref{fig2} shows the spectra
measured for the two PCFs under investigation.

In fiber A (with the ZDWs separated by only 161~nm), we obtained a
relatively narrow spectrum ranging from 1000 nm to about 1200 nm It
is important to note that the upper limit of the SC ($\sim$1200~nm)
is well below the OH absorption peak (1380~nm), which can
consequently not be involved to explain the limitation of the SC
extension. The total average output power was 9~W. In fiber B (with
the ZDWs separated by 285~nm), the SC spectrum is much broader than in
fiber A. It spans from 1050~nm to 1550~nm, with an important dip
around 1400~nm due to important losses at this wavelength due to
water absorption (measured to be about 300~dB/km at 1380~nm). As a
consequence, the average output power is reduced to 6.3~W. Note that
the spectrum generated in fiber B is limited by the pump on the low
wavelength side rather than by the blue-shifted DW. This is due to
the fact that this blue-shifted DW is very weak because its spectral
shift with the soliton is very important \cite{arnaud07}. It is also
important to note that the spectral boundaries of both experimental
spectra are in very good agreement with the ones predicted by
numerical simulations for corresponding PCFs in
Ref.~\cite{arnaud07}. Our experimental results thus clearly reveal
that the SC spectrum is actually bounded by the presence of two DWs
as expected. For both fibers, we checked that the general shape of
the output spectrum and especially its upper and lower limits did
not change when fibers longer than 60~m were used. In that case,
there was only a reduction of the output power due to absorption
losses, and more pump power was converted in the SC. On the
contrary, by shortening the fiber length to less than 60~m, the
bandwidth of the SC was strongly reduced and most of the output
power was located in the vicinity of the pump wavelength. We found
that a length of 60~m was a good compromise between the output power
and the conversion efficiency of the pump power into the SC.

Above the fundamental study and the experimental demonstration
reported in the present paper, we believe that SC-based sources with
a controllable generated spectrum would be of great interest for
numerous applications. However, this would require a very good
temporal stability and an improved flatness of the output spectrum.
Indeed, in the case of the ``free space'' injection setup used in
our study, the output spectrum was stable just over a few minutes at
full pump power. After this duration, the output power and the
stability of the spectrum began to decay because of thermal effects
at the fiber input. Although the stability over a few minutes is
sufficient to record the spectrum with an OSA without any trouble,
it will not be suitable for most of potential applications. This
crucial limitation could be overcome by directly splicing the PCF to
the fiber laser output. A careful splice would also present the
advantage of increasing the amount of pump power launched into the
core by limiting the power coupled into the inner cladding. This
would consequently improve the conversion of the pump power into the
SC, \emph{i.e.} the spectral power density at the fiber output. Finally, the dip around 1400~nm in the spectrum could be removed by reducing the OH contamination with a suitable cleaning
of the stacked PCF preform \cite{travers05}.

In conclusion, we have studied CW SC generation in PCFs with two
zero-dispersion wavelengths located around the pump wavelength. By a
suitable control of the position of both ZDWs and of the dispersion
value at the pump wavelength, we evidenced experimentally for the
fisrt time to our knowledge the possibility of tailoring the SC
spectrum. By this way, the SC extension and the output power can be
adjusted over the spectral range of interest required for a
particular application in the near IR. This represents an important
step towards optimizing the pump power budget for a given
application of SC-based sources. Next steps will consist in
improving the flatness of the SC by reducing the water contamination
and to directly splice the laser source with the PCF. We believe
that this all fiber SC light source would become an useful tool for
OCT applications for example.

The authors acknowledge Karen Delplace (IRCICA) for assistance in
fiber fabrication and Thibault Sylvestre (FEMTO-ST) for fruitful
discussions.

\clearpage

\section*{Table caption}
\begin{table}[h]
\caption{Parameters of the fabricated PCFs.} \label{tab1}
\end{table}
\clearpage

\begin{center}
\begin{tabular}{ccc}\hline
~~~~Parameter~~~~ & ~~~~Fiber A ~~~~& ~~~~Fiber B~~~~ \\
\hline
ZDW1 (nm)&  1010&   976\\
ZDW2 (nm)&  1181&   1261\\
$\gamma$ @ 1064 nm (W$^{-1}$.km$^{-1}$)&    23  &24\\
$D$ @ 1064 nm (ps/nm/km)&   3&  8\\
Length (m)& 60  &60 \\\hline
\end{tabular}
\end{center}
\clearpage

\section*{Figure caption}

\begin{figure}[h]
\caption{GVD curves of the fabricated PCFs calculated with a FEM
package from high resolution SEMs. Inset: SEM of fiber B. The
vertical dashed line represents the pump wavelength (1064~nm).}
\label{fig1}
\end{figure}

\begin{figure}[h]
\caption{(a) Setup used for SC generation experiments. (b)
Experimental spectra recorded for 60~m of fiber A (dashed line) and
of fiber B (plain line) with a pump power of 14~W launched into the
PCFs.} \label{fig2}
\end{figure}

\clearpage

\centering \epsfig{file=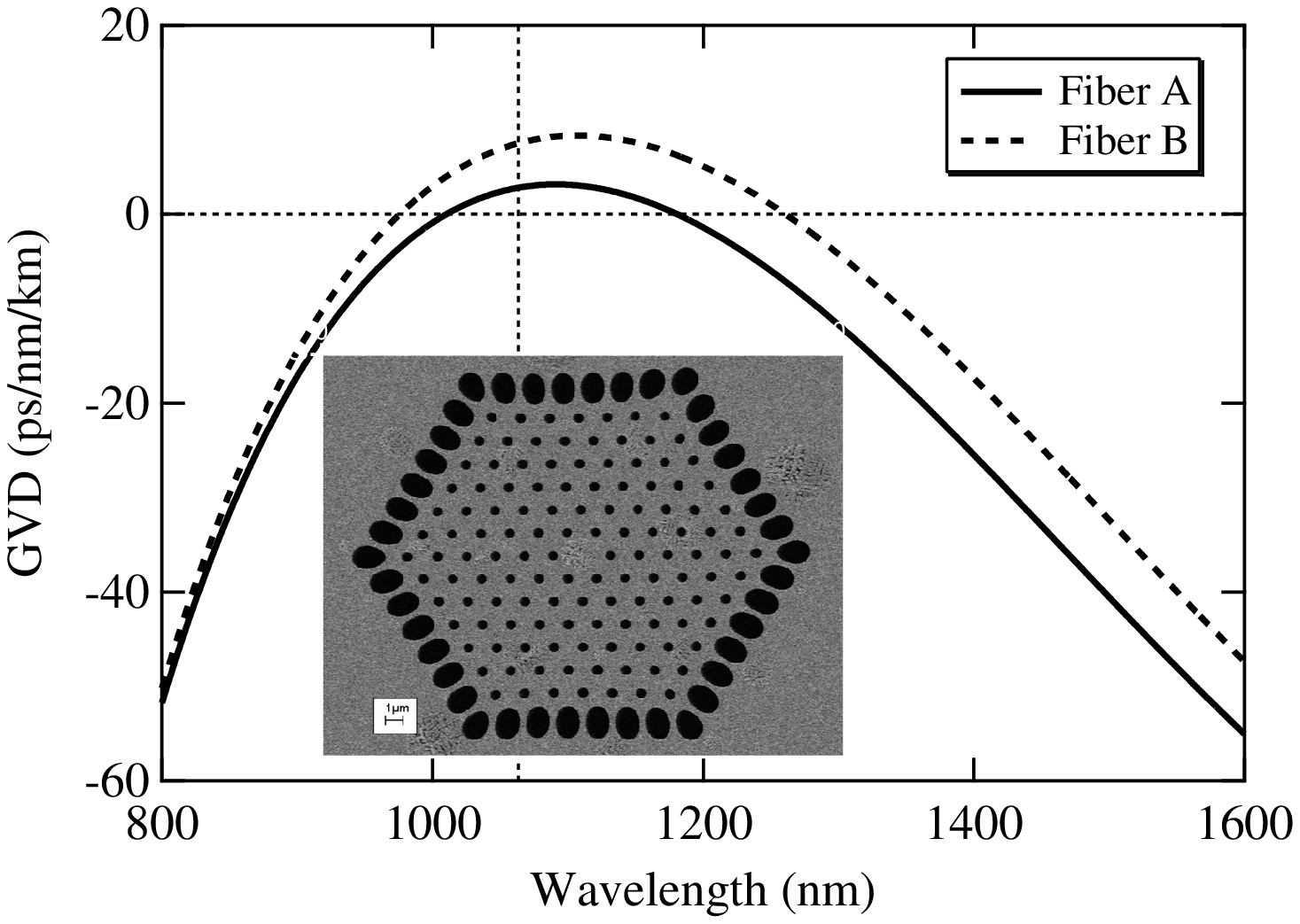,width=8.5cm} \clearpage

\centering \epsfig{file=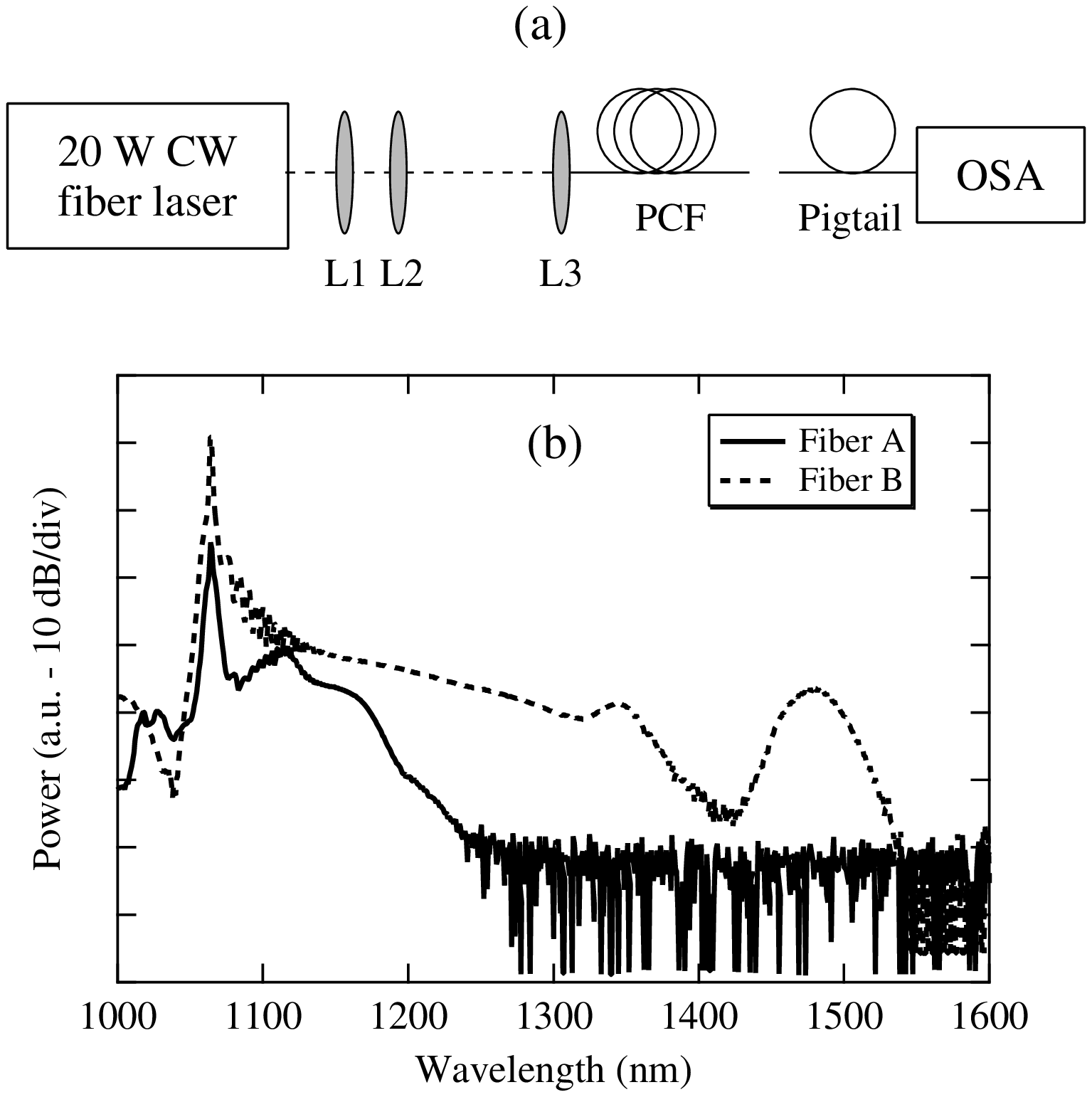,width=8.5cm}
\par\clearpage


\newpage

\begin{thebibliography}{99}
\bibitem{taylor03} A. V. Avdokhin, S. V. Popov, and J. R. Taylor, "Continuous-wave, high-power, Raman continuum generation in holey fibers," \ol \textbf{28}, 1353 (2003).

\bibitem{miguel03} M. Gonz\'{a}lez-Herr\'{a}ez, S. Mart\'{\i}n-L\'{o}pez, P. Corredera, M. L. Hernanz, and P. R. Horche, "Supercontinuum generation using a continuous-wave Raman fiber laser," \oc \textbf{226}, 323 (2003).

\bibitem{abeeluck04} A. K. Abeeluck, C. Headley, and C. G. J{\o}rgensen, "High-power supercontinuum generation in highly nonlinear, dispersion-shifted fibers by use of a continuous-wave Raman fiber laser," \ol \textbf{29}, 2163 (2004).

\bibitem{abeeluck05} A. K. Abeeluck, and C. Headley, "Continuous-wave pumping in the anomalous- and normal-dispersion regimes of nonlinear fibers for supercontinuum generation," \ol \textbf{30}, 61 (2005).

\bibitem{travers05} J. C. Travers, R. E. Kennedy, S. V. Popov, J. R. Taylor, H. Sabert, and B. Mangan, "Extended continuous-wave supercontinuum generation in a low-water-loss holey fiber," \ol \textbf{30}, 1938(2005).

\bibitem{arnaud07}  A. Mussot, M. Beaugeois, M. Bouazaoui, and T. Sylvestre, "Tailoring CW supercontinuum generation in microstructured fibers with two-zero dispersion wavelengths," Opt. Express \textbf{15}, 11553 (2007).

\bibitem{agrawal} G. P. Agrawal, Nonlinear Fiber Optics, 3rd ed., (Academic Press, San Diego, CA, USA, 2001).

\bibitem{cristiani04} I. Cristiani, R. Tedioso, L. Tartara, and V. Degiorgio, "Dispersive wave generation by solitons in microstructured optical fibers," Opt. Express \textbf{12}, 124 (2004).

\bibitem{akhmediev95} N. Akhmediev and M. Karlsson, "Cherenkov radiation emitted by solitons in optical fibers," Phys. Rev. A \textbf{51}, 2602 (1995).

\bibitem{skryabin03} D. V. Skryabin, F. Luan, J. C. Knight, and P. St. J. Russell, "Soliton self-frequency shift cancellation in photonic crystal fibers," Science \textbf{301}, 1705 (2003).

\bibitem{frosz} M. H. Frosz, O. Bang, and A. Bjarklev, "Soliton collision and Raman gain regimes in continuous-wave pumped supercontinuum generation," Opt. Express \textbf{14}, 9391 (2006).

\bibitem{tse06} M. L. V. Tse, P. Horak, F. Poletti, N. G. R. Broderick, J. H. V. Price, J. R. Hayes, and D. J. Richardson, "Supercontinuum generation at 1.06~$\mu$m in holey fibers with dispersion flattened profiles," Opt. Express \textbf{14}, 4445 (2006).

\bibitem{wadsworth04} W. J. Wadsworth, N. Joly, J. C. Knight, T. A. Birks, F. Biancalana, and P. St. J. Russell, "Supercontinuum and four-wave mixing with Q-switched pulses in endlessly single-mode photonic crystal fibers," Opt. Express \textbf{12}, 299 (2004).

\end{thebibliography}
\end{document}